\documentclass[twocolumn,aps,pra]{revtex4}
\usepackage{dcolumn}
\usepackage{graphicx}

\begin{document}

\title{Atomic noise spectra in nonlinear
magneto-optical rotation in a rubidium vapor}


\author{Hebin Li,$^{1}$ Vladimir A. Sautenkov,$^{1,2}$ Tigran S. Varzhapetyan,$^{1,3}$ Yuri V. Rostovtsev,$^1$ and Marlan O. Scully$^{1,4}$}
\address{$^1$Institute for Quantum Studies and Department of Physics,
        Texas A\&M University, College Station, Texas 77843, USA}
\address{$^2$Lebedev Institute of Physics, Moscow 119991, Russia}
\address{$^3$Institute for Physical Research of NAS of Armenia,
Ashtarak-2 378410, Armenia}
\address{$^4$Princeton Institute for the Science and Technology of
Materials and Department of Mechanical \& Aerospace Engineering,
Princeton University, Princeton, New Jersey 08544, USA}

\begin{abstract}
We have studied the noise spectra in a nonlinear magneto-optical
rotation experiment in a rubidium vapor. We observed the reduction
of noise in the intensity difference of two orthogonally polarized
components of the laser beam. The dependence of the noise level on
both the frequency and the longitudinal magnetic field has been
studied. We found that the optimal condition for the noise reduction
is to work around zero longitudinal magnetic field, where the
intensity correlation between the two orthogonally polarized
components is maximum. Our results can be used to reduce or
eliminate the atomic excess noise, therefore improving the
sensitivity of nonlinear magneto-optical rotation magnetometers and
other atom-optical based applications.
\end{abstract}


\maketitle 

\section{INTRODUCTION}

Improving the sensitivity of magnetometers is important both for
practical applications and for fundamental research. Magnetometers
based on atom-optical techniques, such as the optical pumping
magnetometers \cite{Alexandrov1996} and the nonlinear
magneto-optical rotation (NMOR) magnetometers
\cite{Budker2000,Sautenkov2000,Kominis2003}, have achieved
sensitivities of the order of 10$^{-15}$ T Hz$^{-1/2}$. Quantum
noise starts to play a crucial role in obtaining higher sensitivity,
which approaches the atom shot-noise-limited sensitivity
\cite{Budker2002}. One of the contributions to the quantum noise is
the increase of laser beam intensity fluctuation due to the laser
interacting with an atomic vapor (atomic excess noise). Possible
processes that are responsible for the generation of the atomic
excess noise include the conversion of laser phase noise to
intensity noise
\cite{Yabuzaki1991,Mcintyre1993,Camparo1998,Camparo1999} and the
four-wave mixing process \cite{Agarwal1995}. Although the atomic
excess noise could be a useful spectroscopic tool
\cite{Yabuzaki1991,Walser1994}, it is usually not desirable in
atom-optical based applications, such as atom-optical magnetometers,
atomic frequency references \cite{Kitching2001}, and the generation
of squeezed light
\cite{Hetet2007,Matsko2002,Ries2003,Hsu2006,Mikhailov2008}.

To reduce or eliminate the influence of atomic excess noise, one can
take advantage of the intensity correlation properties of the
optical fields passing through an atomic vapor. As is shown in
\cite{Sautenkov2005}, an electromagnetically induced transparency
(EIT) experiment was performed by coupling two beams from one laser
with an excited state and Zeeman sublevels of the ground state in a
rubidium vapor. The authors observed the intensity correlation and
anticorrelation between two circularly polarized laser beams. More
generally, a similar effect was also observed with two beams from
two independent lasers \cite{Cruz2007}. In the case of correlation
or anticorrelation, the intensity noise in each of two laser fields
is fluctuating with a phase difference of 0 or 180$^\circ$. A simple
summation or subtraction of these two signals can suppress the
noise. Experiments have shown the ability of reducing the noise to
the shot-noise level by taking the difference of two laser beams in
an EIT configuration \cite{Sautenkov2007}. In NMOR experiments, the
recent observation of the intensity correlation \cite{Tigran2008},
along with the power spectra study of the noise at 2.5 MHz
\cite{Martinelli2004}, indicate the possibility of using the
intensity correlation to reduce or eliminate the atomic excess
noise.

In this paper, we report the experimental study of the noise spectra
in a nonlinear magneto-optical rotation experiment in a rubidium
vapor. We show that the atomic excess noise in NMOR can be
essentially reduced to the shot-noise level because of the intensity
correlation of two orthogonally polarized components. The dependence
of the noise in the difference signal of these two components on
both frequency and magnetic field, has been studied. Our results
show that the optimal working condition for reducing the atomic
excess noise is to work near zero longitudinal magnetic field.

\section{EXPERIMENTAL SETUP}
\begin{figure}[htb]
\center
\includegraphics[width= 0.9\columnwidth]{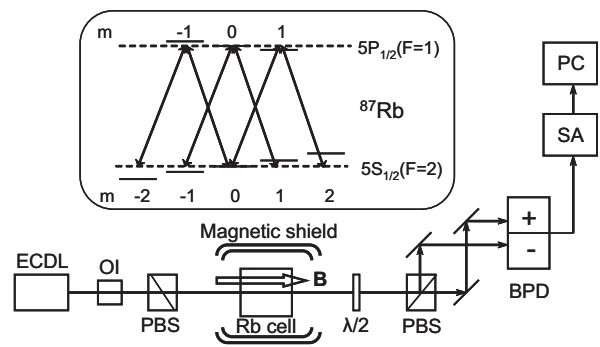}

\caption{\label{1} The experimental schematic and the energy diagram
(the inset). ECDL: external cavity diode laser; OI: optical
isolator; PBS: polarizing beam splitter; $\lambda$/2: half-wave
plate; BPD: balanced photo diode; SA: spectrum analyzer; PC:
computer.}
\end{figure}

The experimental schematic is illustrated in Fig. 1. The laser
source is an external cavity diode laser (ECDL) described in
\cite{Vassiliev2006}. The laser is tuned to the rubidium D$_1$ line
(795 nm), specifically at the transition 5S$_{1/2}$ (F=2)
$\leftrightarrow$ 5P$_{1/2}$ (F=1) of $^{87}$Rb, referenced to the
Doppler-free saturation resonance in a rubidium cell at room
temperature. The frequency drift is less than 30 MHz per hour after
a sufficient warm-up time. The linewidth of the laser emission is
less than 1 MHz. The laser beam has a diameter of 1 mm, and it is
linearly polarized. After passing through an optical isolator, the
beam proceeds through a polarizing beam splitter (PBS) and possesses
a polarization parallel to the optical table.

The beam goes into a glass cell filled with a rubidium vapor that
contains the natural isotope abundance of rubidium atoms. The cell
has the length of 7.5 cm, and it is heated to reach an atomic
density of 10$^{12}$ cm$^{-3}$. A two-layer magnetic shield isolates
the cell from environmental magnetic fields in the lab, while a
solenoid inside the magnetic shield provides an adjustable
longitudinal magnetic field. The linearly polarized beam is a
combination of the left- and right-circularly polarized components.
The two circular components are coupled to the energy levels of
$^{87}$Rb as shown in the energy diagram in Fig. 1.

The output beam from the rubidium cell is analyzed by a half-wave
plate ($\lambda$/2) and a PBS. The half-wave plate is set to rotate
the polarization by 45$^\circ$, such that without the rubidium cell,
the PBS equally splits the intensity of the beam. If the rubidium
cell is placed in the system, a rotation angle of the beam
polarization will be introduced that depends on the magnitude of the
longitudinal magnetic field \cite{Budker2002}. With a nonzero
magnetic field (B$\neq$0), the two beams coming out from the PBS do
not have equal intensities. Recording the intensities of two beams
as $I_1$ and $I_2$, the polarization rotation due to rubidium atoms
can be calculated using the following equation
\begin{equation}
\phi=\arcsin(\frac{I_1-I_2}{I_1+I_2})\ .
\end{equation}
To study the power spectra of the atomic excess noise, a balanced
photo detector (BPD) with a sensitivity of 2$\times$10$^4$ V/W and a
bandwidth from DC to 100 MHz is used to register the intensities of
two laser beams. The optical path lengths of the beams going into
two channels of the BPD are chosen to be the same so that no
additional time delay between the two channels is introduced. The
signal is analyzed by an RF spectrum analyzer. In the case of a zero
magnetic field (B=0), for example, each channel of the BPD records
an intensity
\begin{equation}
I_i=I_0+I(t)+\delta I_i(t)\ ,\ (i=1,2)\ ,
\end{equation}
where $I_0$ is the average intensity, $I(t)$ is the low frequency
intensity fluctuations, and $\delta I_i(t)$ is the atomic excess
noise. Then, the difference signal $\Delta I$ from the BPD is given
by $\Delta I(t) =\delta I_1(t)-\delta I_2(t)$. The spectrum analyzer
gives the Fourier transform of the time dependence of the signal.

\section{EXPERIMENTAL RESULTS AND DISCUSSION}

\begin{figure}[hbt]
\includegraphics[width= 0.47\columnwidth]{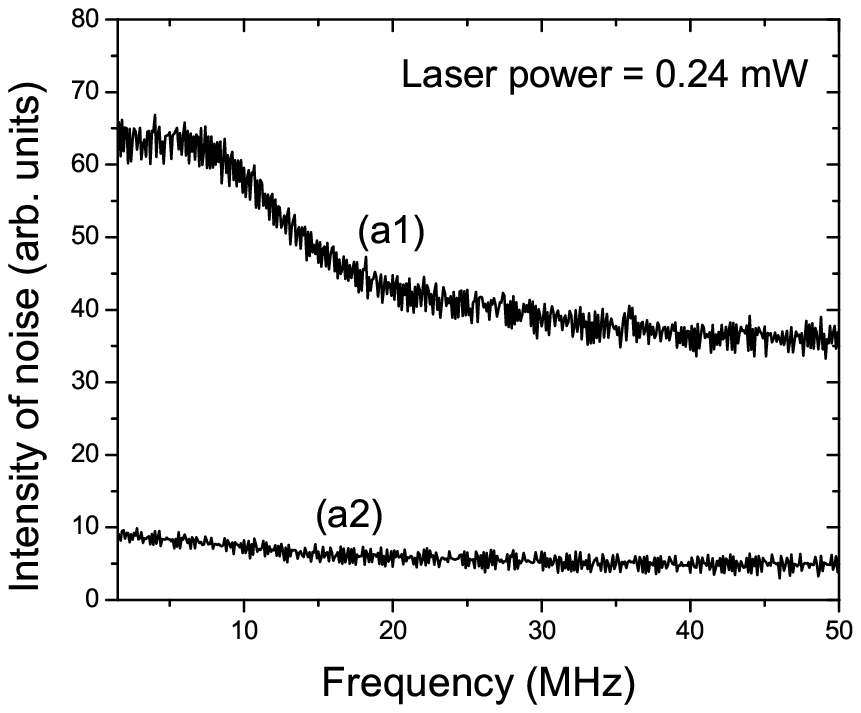}
\includegraphics[width= 0.47\columnwidth]{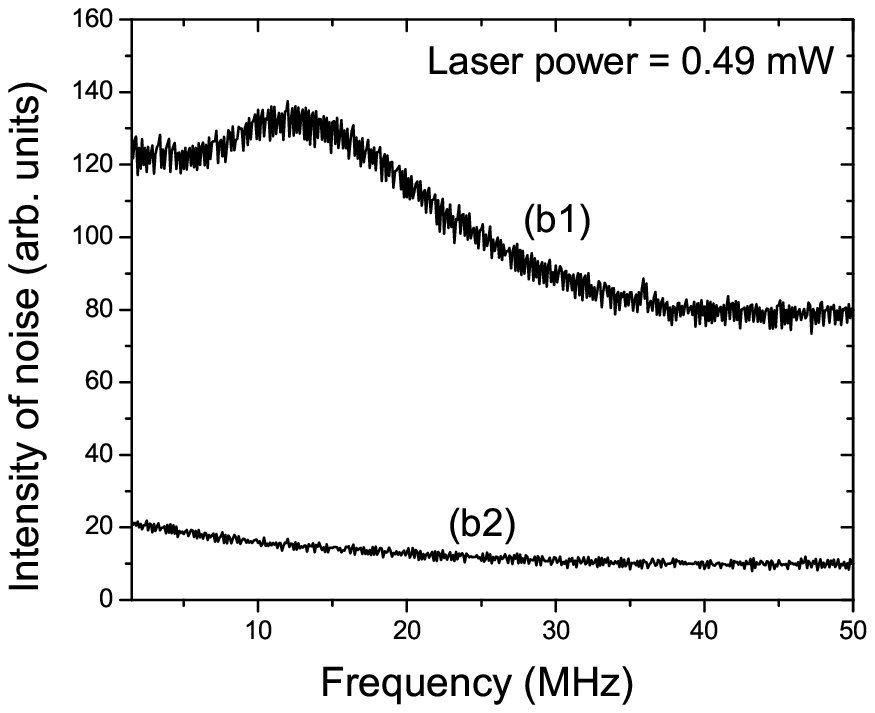}

\caption{\label{2} The power spectra of the noise from laser beams
in an NMOR experiment with the magnetic field B=0. The left and
right figures show the spectra with input laser powers of 0.24 mW
and 0.49 mW, respectively. Traces (a1) and (b1) are the noise
spectra of one laser beam. Traces (a2) and (b2) are the noise
spectra of the balanced signal (with both beams). The spectrum
analyzer was setup with a resolution of 300 kHz and a video
bandwidth of 100 Hz.}
\end{figure}

We begin the presentation of the results by showing the power
spectra of noise with no magnetic field (B=0). Figure 2 shows the
noise spectra for different input laser intensities. The left and
right figures display the spectra for input laser powers of 0.24 mW
and 0.49 mW, respectively. Traces (a2) and (b2) are recorded with
two laser beams sent to the BPD, and they show the noise spectra of
the difference signal. The noise is larger in the low frequency
region. The noise level approaches the shot-noise level
\cite{Sautenkov2007} at higher frequencies. For comparison, traces
(a1) and (b1) are recorded with only one laser beam sent to the BPD.
They represent the noise spectra of the laser beam passing through
the rubidium vapor. Before entering the cell, the laser beam has
small intensity noise but large phase noise. The phase noise is
converted into the intensity noise due to the laser interacting with
the atoms. This process causes a substantial increase of intensity
fluctuations in the laser beam coming out of the cell
\cite{Yabuzaki1991}. Our results show that these intensity
fluctuations can be suppressed by subtracting the intensity of one
laser beam from the other. Comparing the noise spectra of one laser
beam and of the difference signal, (a1) and (a2) for instance, the
noise level of the difference signal is dramatically reduced. When
the input laser power is doubled (0.49 mW), the corresponding
spectra presented as (b1) and (b2) show the same behavior, although
the shot-noise level increases approximately two times because of
the higher laser power. These results can be understood as a
consequence of the intensity correlation between the two output
laser beams from the PBS. As is shown in \cite{Tigran2008}, the
intensities of two beams in an NMOR experiment are highly correlated
(correlation function $G^{(2)}(0)\approx 0.9$) at zero magnetic
field (B=0). The fluctuations $\delta I_1(t)$ and $\delta I_2(t)$
are varying simultaneously, and thus $\Delta I(t)$ will be small.

\begin{figure}[htb]
\center
\includegraphics[width= 0.8\columnwidth]{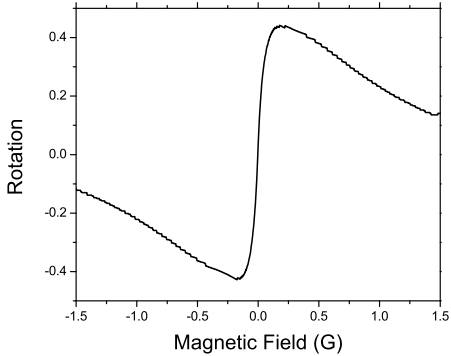}
\caption{\label{1} The polarization rotation is plotted as a
function of the longitudinal magnetic field B.}
\end{figure}

The intensity correlation and the substantial reduction of noise in
NMOR experiments is not trivial in terms of various behaviors at
different magnitudes of the magnetic field B. To show this, besides
the preceding results with a zero magnetic field, we have also
studied the noise spectra in NMOR at magnetic fields of various
magnitudes. Prior to showing these results, a typical measurement of
the polarization rotation in our experiment (laser power P=0.24 mW)
is presented in Fig. 3, to remind us of the rotation dependence on
the magnetic field.

\begin{figure}[htb]
\center
\mbox{\includegraphics[width= 0.47\columnwidth]{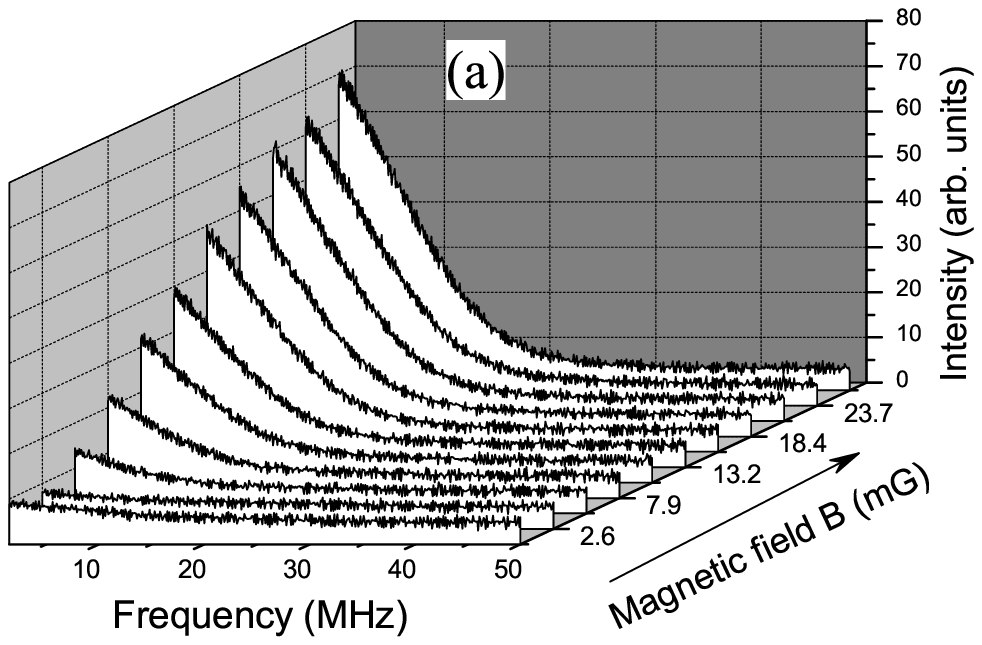}}
\mbox{\includegraphics[width= 0.47\columnwidth]{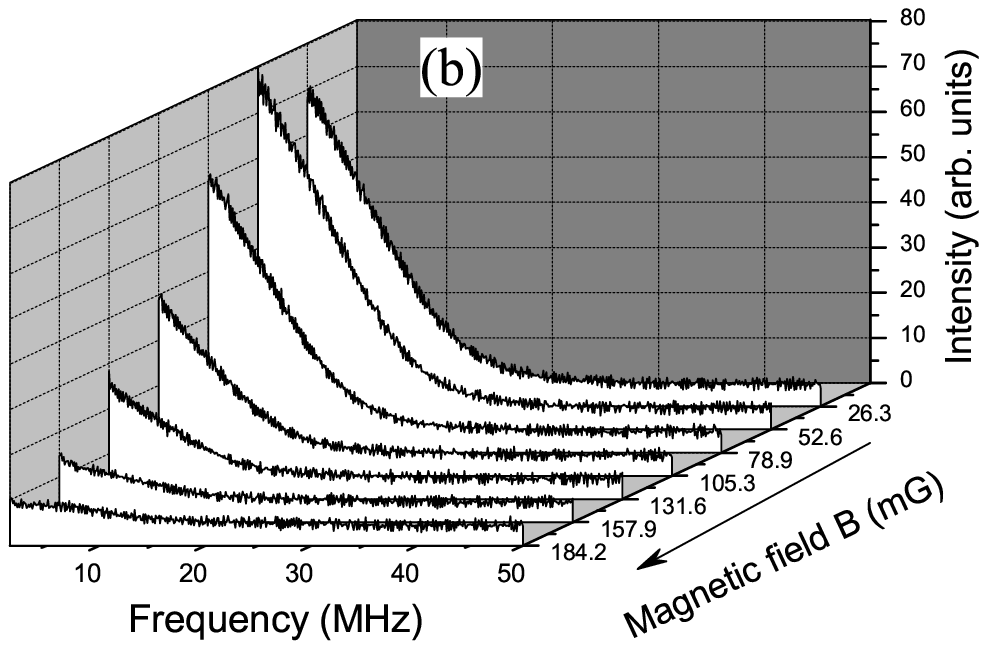}}
\caption{\label{1} The noise spectra dependence on the longitudinal
magnetic field. The noise level is plotted as a function of both the
frequency and the magnetic field. (a) The spectra corresponding to
magnetic fields ranging from 0 to 26.3 mG; (b) The the spectra
corresponding to magnetic fields ranging from 26.3 mG to 184 mG. The
arrows denote the ascending direction of the magnitude of the
magnetic field.}
\end{figure}

We record the noise spectra of the difference signal with two output
beams sent to the BPD for several magnetic fields. The results are
presented as three dimensional plots in Fig. 4 (laser power P=0.24
mW). The magnetic field varies from 0 to 184 mG. The spectra
corresponding to magnetic fields ranging from 0 to 26.3 mG are shown
in plot (a), and the ones corresponding to magnetic field ranging
from 26.3 mG to 184 mG are shown in plot (b). The spectra are sorted
by the magnetic field ascending along the arrows shown in the
figure. The magnetic field in plot (a) steps by about 2.6 mG, while
it steps by about 26 mG in plot (b). Note that the noise spectra
corresponding to negative magnetic fields, which are not plotted
here, have the symmetric behaviors. From these results, we see that
the reduction of high frequency noise is nearly the same for
different magnetic fields, but the low frequency noise is not
appreciably reduced. This shows that the low frequency noise is not
correlated at all magnitudes of the magnetic field, but the high
frequency noise is better correlated. A detailed study of the noise
spectra dependence on the magnetic field for each individual NMOR
system can provide a guideline for choosing optimal working
parameters to reduce or eliminate the atomic excess noise.

\begin{figure}[htb]
\center
\includegraphics[width= 0.49\columnwidth]{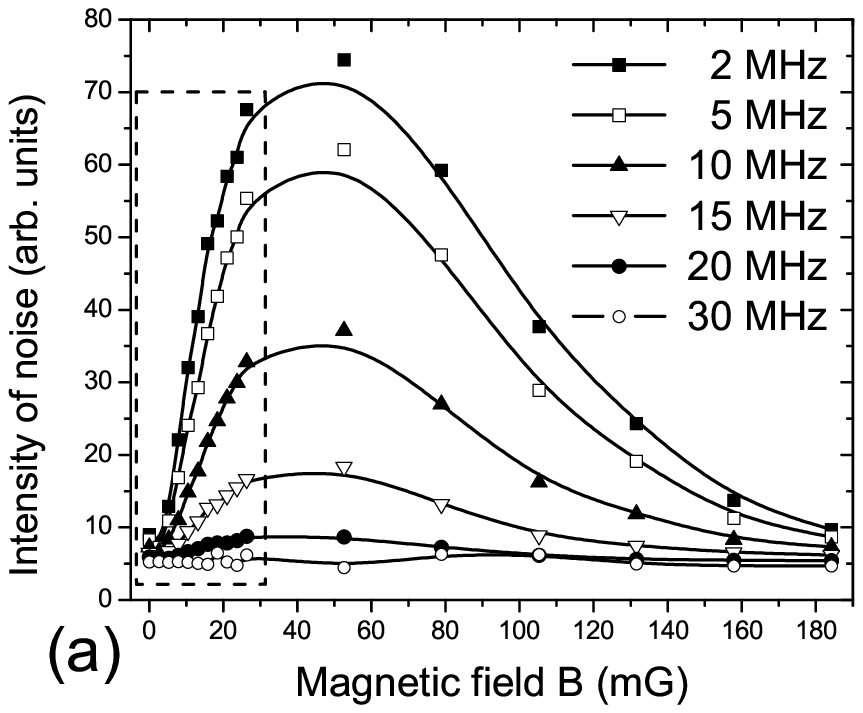}
\includegraphics[width= 0.49\columnwidth]{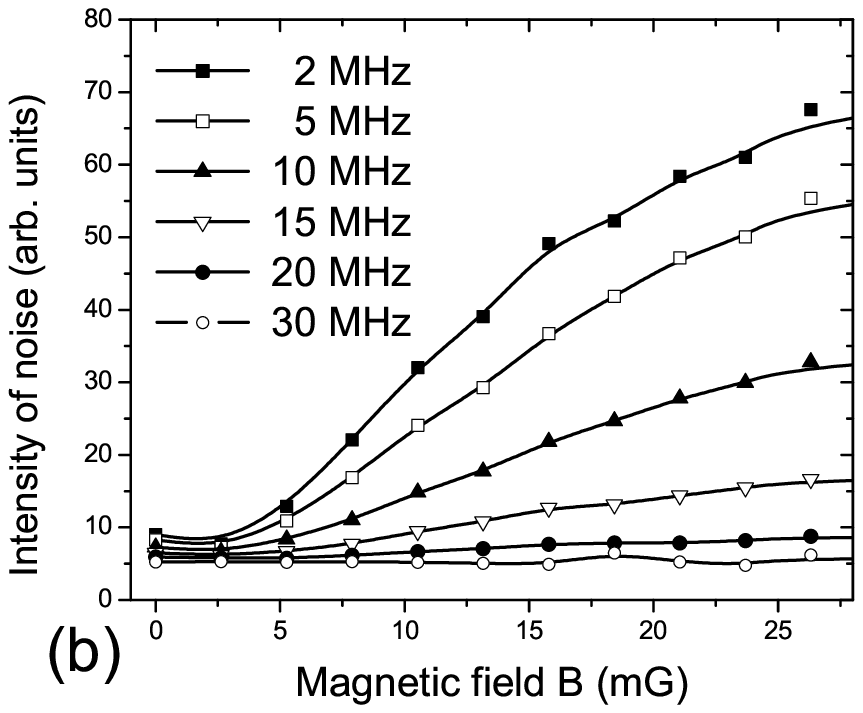}
\caption{\label{1} (a) The level of noise at different frequencies
is plotted as a function of the magnitude of the magnetic field. (b)
The magnification of the dashed square region in (a). Different
symbols denote different frequencies. Square: 2 MHz; hollow square:
5 MHz; triangle: 10 MHz; hollow triangle: 15 MHz; dot: 20 MHz;
circle: 30 MHz. The solid lines are smooth connections of the data
points.}
\end{figure}

To better demonstrate how the noise at a certain frequency depends
on the magnitude of the magnetic field, we cut the three dimensional
plots in Fig. 4 at specific frequencies along the plane of the
magnetic field axis and the intensity axis. The cross sections
picked up are shown in Fig. 5(a), in which the symbols square,
hollow square, triangle, hollow triangle, dot and circle represent
the level of noise at frequencies of 2 MHz, 5 MHz, 10 MHz, 15 MHz,
20 MHz and 30 MHz, respectively. At the zero magnetic field, the
noise is suppressed to the shot-noise level over the entire
frequency region. For the frequencies lower than 20 MHz, the noise
level increases quickly within about 50 mG, and comes back close to
the shot-noise level within about 130 mG. For the frequencies higher
than 20 MHz, the noise level essentially remains close to the
shot-noise level. To show the details of the rising slope in the
dashed square in Fig. 5(a), a magnification of this region is shown
in Fig. 5(b). For low frequency noise, the noise level remains
nearly minimum value not only at the zero magnetic field but also in
a small region around zero magnetic field. In our data, the noise
level of 2 MHz and 5 MHz noise is almost flat for magnetic fields
ranging from 0 to 2.5 mG.

These results show that, for the best reduction of atomic excess
noise in atom-optical applications, one should work near a zero
longitudinal magnetic field. However, a rigorously exact zero
magnetic field is not necessary, because the same reduction of noise
can be obtained in a region around zero magnetic field. This
property makes the implementation relatively easier and more
reliable. As an example, to obtain the best reduction of the atomic
excess noise in NMOR magnetometers, one might use an external
calibrated magnetic field to compensate \cite{Budker2002}, so that
longitudinal magnetic field is close to zero.

\section{CONCLUSION}

We have experimentally studied the noise spectra in a nonlinear
magneto-optical rotation experiment in rubidium vapor. We have shown
that a detailed study of noise reduction, due to the intensity
correlation between two orthogonally polarized components of the
laser beam, can suggest the optimal working conditions for reducing
atomic excess noise. The noise in the difference signal of two
orthogonal components at different frequencies has been studies as a
function of magnetic field. The study of the noise dependence on
both the noise frequency and the magnetic field shows that the
maximum reduction of noise can be obtained around zero longitudinal
magnetic field.

Our results can be used to reduce or eliminate atomic excess noise,
and thus improve the sensitivity of NOMR magnetometers. The study
also indicates the potential importance of the intensity
correlations in other atom-optical applications such as atomic
frequency references and the generation of squeezed light.

\section{ACKNOWLEDGEMENT}
We thank D. Budker, M.M. Kash, E.E. Mikhailov, F.A. Narducci, I.
Novikova, D. Sarkisyan, G.R. Welch, and M.S. Zubairy for useful and
fruitful discussions, and gratefully acknowledge the support from
the Office of Naval Research, the Robert A.\ Welch Foundation (Grant
\#A1261). One of us (T.S.V) also thanks NFSAT\ award TFP 2005/02 for
financial support and the IQS and the Department of Physics of Texas
A\&M University for their hospitality.


\begin{thebibliography}{99}

\bibitem{Alexandrov1996} E.B. Alexandrov, M.V. Balabas, A.S. Pasgalev, A.K. Vershovskii, and N.N.
Yakobson, ``Double-resonance atomic magnetometers: From gas
discharge to laser pumping," Laser Phys. \textbf{6}, 244 (1996).

\bibitem{Budker2000} D. Budker, D.F. Kimball, S.M. Rochester, V.V.
Yashchuk, and M. Zolotorev, ``Sensitive magnetometry based on
nonlinear magneto-optical rotation," Phys. Rev. A \textbf{62},
043403 (2000).

\bibitem{Sautenkov2000} V.A. Sautenkov, M.D. Lukin, C.J. Bednar, I,
Novikova, E. Mikhailov, M. Fleischhauer, V.L. Velichansky, G.R.
Welch, and M.O. Scully, ``Enhancement of magneto-optic effects via
large atomic coherence in optically dense media," Phys. Rev. A
\textbf{62}, 023810 (2000).

\bibitem{Kominis2003} I. K. Kominis, T. W. Kornack, J. C. Allred, and M. V.
Romalis, ``A subfemtotesla multichannel atomic magnetometer," Nature
\textbf{422}, 596 (2003).

\bibitem{Budker2002} D. Budker, W. Gawlik, D.F. Kimball, S.M. Rochester, V. V. Yashchuk, and A.
Weis, ``Resonant nonlinear magneto-optical effects in atoms," Rev.
Mod. Phys. \textbf{74}, 1153 (2002).

\bibitem{Yabuzaki1991} T. Yabuzaki, T. Mitsui, and U. Tanaka, ``New type of high resolution
sepctroscopy with a diode laser," Phys. Rev. Lett. \textbf{67}, 2453
(1991).

\bibitem{Mcintyre1993} D.H. McIntyre, C.E. Fairchild, J. Cooper, and
R. Walser, ``Diode-laser noise spectroscopy of rubidium," Opt. Lett.
\textbf{18}, 1816 (1993).

\bibitem{Camparo1998} J.C. Camparo, ``Conversion of laser phase noise to amplitude
noise in an optically thick vapor," J. Opt. Soc. Am. B \textbf{15},
1177 (1998).

\bibitem{Camparo1999} J.C. Camparo and J.G. Coffer, ``Conversion of laser phase noise to
amplitude noise in a resonant atomic vapor: The role of laser
linewidth," Phys. Rev. A \textbf{59}, 728 (1999).

\bibitem{Agarwal1995} R. J. Gehr, A. L. Gaeta, and R. W. Boyd, and G.S.Agarwal, ``Excess noise acquired
by a laser beam after propagating through an atomic-postassium
vapor," Phys. Rev. A \textbf{51}, 4152 (1995).

\bibitem{Walser1994} R. Walser and P. Zoller, ``Laser-noise-induced
polarization fluctuation as a spectroscopic tool," Phys. Rev. A
\textbf{49}, 5067 (1994).

\bibitem{Kitching2001} J. Kitching, H.G. Robinson, L. Hollberg, S.
Knappe, and R. Wynands, ``Optical-pumping noise in laser-pumped,
all-optical microwave frequency references," J. Opt. Soc. Am. B
\textbf{18}, 1676 (2001).

\bibitem{Hetet2007} G. H\'{e}tet, O. Gl\"{o}ckl, K.A. Pilypas, C.C. Harb, B.C.
Buchler, H.-A. Bachor, and P.K. Lam, ``Squeezed light for
bandwidth-limited atom optics experiments at the rubidium D1 line,"
Journal of Physics B \textbf{40}, 221 (2007).

\bibitem{Matsko2002} A.B. Matsko, I. Novikova, G.R. Welch, D. Budker,
D.F. Kimball, and S.M. Rochester, ``Vacuum squeezing in atomic media
via self-rotation," Phys. Rev. A \textbf{66}, 043815 (2002).

\bibitem{Ries2003} J. Ries, B. Brezger, and A.I. Lvovsky, ``Experimental
vacuum squeezing in rubidium vapor via selfrotation," Phys. Rev. A
\textbf{68}, 025801 (2003).

\bibitem{Hsu2006} M.T.L. Hsu, G. H\'{e}tet, et. al. ``Effect of
atomic noise on optical squeezing via polariztion self-rotation in a
thermal vapor cell," Phys. Rev. A \textbf{73}, 023806 (2006).

\bibitem{Mikhailov2008} E.E. Mikhailov and I. Novikova,
``Low-frequency vacuum squeezing via polarization self-rotation in
Rb vapor," arXiv:0802.1558v1 (2008).

\bibitem{Sautenkov2005} V.A. Sautenkov, Y.V. Rostovtsev, and M.O.
Scully, ``Switching between photon-photon correlations and Raman
anticorrelations in a coherently prepared Rb vapor," Phys. Rev. A
\textbf{72}, 065801 (2005).

\bibitem{Cruz2007} L.S. Cruz, D. Felinto, J.G. Aguirre G\'{o}mez,
M. Martinelli, P. Valente, A. Lezama, and P. Nussenzveig,
``Laser-noise-induced correlations and anti-correlations in
electromagnetically induced transparency," Eur. Phys. J. D
\textbf{41}, 531 (2007).

\bibitem{Sautenkov2007} V.A. Sautenkov, H. Li, Y.V. Rostovtsev, and M.O.
Scully, ``Power spectra and correlations of intensity fluctuations
in electromagnetically induced transparency," J. Mod. Opt.
\textbf{54}, 2451 (2007).

\bibitem{Tigran2008} T.S. Varzhapetyan, H. Li, G.O. Ariunbold, V.A.
Sautenkov, Y.V. Rostovtsev, and M.O. Scully, ``Intensity
correlations in resonance nonlinear magneto-opticcal rotation,"
arXiv:0803.3050v1 (2008).

\bibitem{Martinelli2004} M. Martinelli, P. Valente, H. Failache, D.
Felinto, L.S. Cruz, P. Nussenzveig, and A. Lezama, ``Noise
spectroscopy of nonlinear magneto-optical resonances in Rb vapor,"
Phys. Rev. A \textbf{69}, 043809 (2004).

\bibitem{Vassiliev2006} V.V. Vassiliev, S.A. Zibrov and V.L.
Velichansky, ``Compact extended-cavity diode laser for atomic
spectroscopy and metrology" Rev. Sci. Instrum. \textbf{77}, 013102
(2006).
\end{thebibliography}
\end{document}